\documentstyle[12pt]{article}
\def\nc{\newcommand}
\def\nn{\nonumber}
\def\p{\partial}

\nc{\bA}{\mbox{\boldmath $A$\unboldmath}}
\nc{\bn}{\mbox{\boldmath $n$\unboldmath}}
\nc{\bl}{\mbox{\boldmath $l$\unboldmath}}
\nc{\bm}{\mbox{\boldmath $m$\unboldmath}}

\nc{\pr}{\frac{\p}{\p r}}
\nc{\pv}{\frac{\p}{\p v}}
\nc{\pta}{\frac{\p}{\p \theta}}
\nc{\pvi}{\frac{\p}{\p \varphi}}

\nc{\pdr}{\frac{\p^2}{\p r^2}}
\nc{\pdv}{\frac{\p^2}{\p v^2}}
\nc{\pdta}{\frac{\p^2}{\p \theta^2}}
\nc{\pdvi}{\frac{\p^2}{\p \varphi^2}}
\nc{\pdva}{\frac{\p^2}{\p v \p \varphi}}
\nc{\pdvr}{\frac{\p^2}{\p v \p r}}
\nc{\pdri}{\frac{\p^2}{\p r \p \varphi}}

\nc{\spr}{\frac{\p}{\p r_*}}
\nc{\spv}{\frac{\p}{\p v_*}}
\nc{\spta}{\frac{\p}{\p \theta_*}}

\nc{\spdr}{\frac{\p^2}{\p r_*^2}}
\nc{\spdvr}{\frac{\p^2}{\p r_* \p v_*}}
\nc{\spdra}{\frac{\p^2}{\p r_* \p \theta_*}}
\nc{\spdri}{\frac{\p^2}{\p r_* \p \varphi}}

\nc{\sta}{\sin\theta}
\nc{\cta}{\cos\theta}
\nc{\sda}{\sin^2\theta}
\nc{\cda}{\cos^2\theta}
\nc{\coa}{\cot\theta}
\nc{\sqd}{\sqrt{2}}
\nc{\cD}{\cal D}
\nc{\cL}{\cal L}
\nc{\cLd}{{\cal L}^{\dagger}}
\nc{\drH}{\dot{r}_H}
\nc{\ddrH}{\ddot{r}_H}
\nc{\prH}{r_H^{\prime}}
\nc{\ka}{\kappa}
\nc{\tkr}{2\ka(r-r_H)}
\nc{\pprH}{r_H^{\prime\prime}}

\begin{document}
\pagestyle{myheadings}
\markboth{}{Hawking Radiation of Photons in
$\cdots$ \hfill Wu and Cai~~~~}

\title{\bf Hawking Radiation of Photons in \\
a Vaidya-de Sitter Black Hole}

\author{S. Q. Wu\thanks{E-mail: sqwu@iopp.ccnu.edu.cn} and
X. Cai\thanks{E-mail: xcai@ccnu.edu.cn}\\
\small\it Institute of Particle Physics, Hua-Zhong \\
\small\it Normal University, Wuhan 430079, P.R. China}
\date{}
\maketitle

\baselineskip 20pt
\begin{abstract}
Hawking evaporation of photons in a Vaidya-de Sitter black hole
is investigated by using the method of generalized tortoise
coordinate transformation. Both the location and the temperature
of the event horizon depend on the time. It is shown that Hawking
radiation of photons exists only for the complex Maxwell scalar
$\phi_0$ in the advanced Eddington-Finkelstein coordinate system.
This asymmetry of Hawking radiation for different components of
Maxwell fields probably arises from the asymmetry of spacetime
in the advanced Eddington-Finkelstein coordinate system. It is shown
that the black body radiant spectrum of photons resembles that of
Klein-Gordon particles.


\end{abstract}

\newpage
\section{Introduction}

Hawking's investigation of quantum effects \cite{Hawk} interpreted as the
emission of a thermal spectrum of particles by a black hole event horizon
sets a significant landmark on black hole physics. In the last few decades,
much work has been done on the Hawking evaporation of black holes in some
spherically symmetric and non-static spacetimes \cite{KCKY,Zhaoel}. In a
recent paper \cite{WC} (here refer to Paper I), we re-examined the Hawking
effect of Dirac particles in a Vaidya-type black hole by means of the
generalized tortoise transformation method. We considered simultaneously
the asymptotic behaviors of the first-order and second-order forms of Dirac
equations near the event horizon, and eliminated the crossing-terms of the
first-order derivatives in the second-order equations by using the relations
between the first-order derivatives of the radial Dirac equations. We showed
that the Hawking radiation exists only for $P_2, Q_1$ components of Dirac
spinors in virtue of the restricts imposed by the limiting form of its
first-order equations. We conceived that this asymmetry of the Hawking
radiation for different spinorial components probably originates from the
asymmetry of space-time in the advanced Eddington-Finkelstein coordinate
system.

In this paper, we deal with the thermal radiation of photons in a Vaidya-de
Sitter space-time. The method used here is the same as that presented in
Paper I, namely, we consider the asymptotic behaviors of the first-order
and second-order forms of Maxwell equations in the vicinity of the event
horizon, and recast each second-order equation to a standard wave equation
near the event horizon. The location and the temperature of the event
horizon are shown to be dependent on the time. The black body radiation
spectrum of photons resembles that of Klein-Gordon scalar particles. We
find that due to the restricts put on the Hawking radiation by the limiting
form of the first-order Maxwell equations, not all Maxwell complex scalars
but $\phi_0$ displays the property of thermal radiation. This asymmetry of
Hawking radiation for different field components in a Vaidya-type spacetime
is thought to be a common feature shared by all particles with higher-spins.

The paper is organized as follows: In Sec. 2, the explicit form of sourceless
Maxwell equations within the framework of Newman-Penrose formalism \cite{NP}
are written out by choosing an appropriate null tetrad in the Vaidya-de Sitter
geometry. By using the method of generalized tortoise coordinate transformation,
the event horizon equation is extracted in Sec. 3 from the asymptotic forms of
the radial parts of the first-order Maxwell equations near the event horizon.
Then the second-order radial equations are manipulated in Sec. 4 by the same
procedure and recast into a standard wave equation near the event horizon, in
the meanwhile, an exact expression of the \lq\lq surface gravity" of the event
horizon is also obtained by adjusting the parameter $\ka$. Sec. 5 is devoted
to deriving the thermal radiation spectrum of photons from the event horizon.
Finally we present some discussions in Sec. 6.

\section{Sourceless Maxwell Equations}

The metric of a Vaidya black hole with a cosmological constant $\Lambda$
is given in the advanced Eddington-Finkelstein coordinate system by
\begin{equation}
ds^2 = 2dv(G dv -dr) -r^2(d\theta^2 +\sin^2\theta d\varphi^2) \, ,
\end{equation}
where $2G = 1 -2M/r -\Lambda r^4/3$, in which mass $M(v)$ of the hole
is a function of the advanced time $v$.

The geometry of this Petrov type-D space-time is characterized by two
kinds of surfaces of particular interest: the apparent horizon $r_{AH}
= 2M$ (coincides with the timelike limit surfaces $r_{TLS}$) and the
event horizon $r_{EH} = r_H$. The apparent horizon is the out-most
trapped surface, while the event horizon is necessary a null-surface
$r = r(v)$ that satisfies the null-surface conditions $g^{ij}\p_i F
\p_j F = 0$ and $F(v,r) = 0$. Traditionally the latter is calculated
approximately by the simple physical condition that the photons at the
event horizon are stuck or unaccelerated in the sense that $\ddot{r}
\approx 0$. An effective method to determine the location of the event
horizon of a dynamic black hole is called as generalized tortoise
coordinate transformation (GTCT) with which we can apply it to the
null hypersurface equation $g^{ij}\p_i F\p_j F = 0$ and then take the
limits approaching the event horizon. However in Paper I, the event
horizon equation is extracted from the asymptotic forms of the radial
parts of the first-order Dirac equations near the event horizon. In
this paper, we derive it by the same procedure but with the Maxwell
equations in place of the Dirac equation here.

When the back reaction of the massless spin-$1$ test particles on the
background geometry is neglected, the electromagnetic field equation
is given by the Maxwell equation on a fixed spacetime (1). In order to
write out its explicit form in the Newman-Penrose \cite{NP} formalism,
we establish the following complex null tetrad system $\{\bl, \bn, \bm,
\overline{\bm}\}$ that satisfies the orthogonal conditions $\bl \cdot
\bn = -\bm \cdot \overline{\bm} = 1$. Thus the covariant one-forms can
be written as
\begin{eqnarray}
&&\bl = dv \, , ~~~~~~~~~~~~ \bm = \frac{-r}{\sqd}\Big(d\theta
+i\sta d\varphi\Big) \, ,\nn \\
&&\bn = G dv -dr \, , ~~\overline{\bm} =
\frac{-r}{\sqd}\big(d\theta -i\sta d\varphi\Big) \, ,
\end{eqnarray}
and their corresponding directional derivatives are
\begin{eqnarray}
&&D = -\pr \, , ~~~~~~~~~~ \delta = \frac{1}{\sqd r}\Big(\pta
+\frac{i}{\sta}\pvi\Big) \, , \nn\\
&&\Delta = \pv +G\pr \, , ~~ \overline{\delta} =
\frac{1}{\sqd r}\Big(\pta -\frac{i}{\sta}\pvi\Big) \, .
\end{eqnarray}
It is not difficult to determine the non-vanishing Newman-Penrose
complex spin coefficients \cite{NP} in the above null-tetrad as
\begin{eqnarray}
\rho = 1/r \, , &&\gamma = -G_{,r}/2 = -dG/2dr \, ,\nn\\
\mu = G/r \, , &&\beta = -\alpha = \coa/(2\sqd r)  \, .
\end{eqnarray}

Inserting for the appropriate spin coefficients into the sourceless
Maxwell equations \cite{NP,CC} in the Newman-Penrose formalism \cite{NP}
\begin{eqnarray}
&&(D -2\rho)\phi_1 -(\overline{\delta} +\tilde{\pi}
-2\alpha)\phi_0 = -\tilde{\kappa}\phi_2 \, , \nn\\
&&(\delta -2\tau)\phi_1 -(\Delta +\mu -2\gamma)\phi_0
 = -\sigma\phi_2 \, , \nn\\
&&(D +2\epsilon -\rho)\phi_2 -(\overline{\delta} +2\tilde{\pi})
\phi_1 = -\tilde{\lambda}\phi_0 \, , \nn\\
&&(\delta +2\beta -\tau)\phi_2 -(\Delta +2\mu)\phi_1
= -\tilde{\nu}\phi_0 \, ,
\end{eqnarray}
we obtain
\begin{eqnarray}
&&\Big(\pr +\frac{2}{r}\Big)\phi_1
+\frac{1}{\sqd r}{\cL}_1\phi_0 = 0 \, , \nn \\
&&\Big({\cD} +\frac{G}{r} +G_{,r} \Big)\phi_0
-\frac{1}{\sqd r}{\cLd}_0\phi_1 = 0 \, , \nn \\
&&\Big(\pr +\frac{1}{r}\Big)\phi_2
+\frac{1}{\sqd\rho}{\cL}_0 \phi_1 = 0 \, , \nn \\
&&\Big({\cD} +\frac{2G}{r} \Big)\phi_1
-\frac{1}{\sqd\rho^*}{\cLd}_1\phi_2 = 0 \, , \label{ME}
\end{eqnarray}
here we have defined operators
\begin{eqnarray*}
&&{\cD} = \pv +G\pr \, , \\
&&{\cL}_n=\pta +n\coa -\frac{i}{\sta}\pvi \, , \\
&&{\cLd}_n=\pta +n\coa +\frac{i}{\sta}\pvi \, .
\end{eqnarray*}
By substituting $\Phi_0 = r\phi_0$, $\Phi_1 = \sqd r^2\phi_1$,
$\Phi_2 = r\phi_2$ into Eq. (\ref{ME}), we have
\begin{eqnarray}
&&\pr\Phi_1 +{\cL}_1\Phi_0 = 0 \, ,
~~~~~~ 2r^2\Big({\cD} +G_{,r}\Big)\Phi_0
-{\cLd}_0\Phi_1 = 0 \, , \nn \\
&&2r^2\pr\Phi_2 +{\cL}_0\Phi_1 = 0 \, ,
~~~~~~~~~~~~~~~~~ {\cD}\Phi_1 -{\cLd}_1\Phi_2 = 0 \, .
\label{MP}
\end{eqnarray}

\section{Event Horizon}

Eq. (\ref{MP}) can be decoupled as
$$\Phi_0 = R_0(v,r)S_0(\theta,\varphi) \, ,
~~\Phi_1 = R_1(v,r)S_1(\theta,\varphi) \, ,
~~\Phi_2 = R_2(v,r)S_2(\theta,\varphi)$$
to the radial part
\begin{eqnarray}
&&\pr R_1 +\lambda R_0 = 0 \, ,
~~~~~~ 2r^2\Big({\cD} +G_{,r}\Big)R_0 +\lambda R_1 = 0 \, , \nn\\
&&2r^2\pr R_2 +\lambda R_1 = 0 \, ,
~~~~~~~~~~~~~~~~~ {\cD} R_1 +\lambda R_2 = 0 \, , \label{sepa}
\end{eqnarray}
and the angular part
\begin{eqnarray}
&&{\cL}_1 S_0 = \lambda S_1 \, ,
~~~~~{\cLd}_0 S_1 = -\lambda S_0 \, , \nn\\
&&{\cL}_0 S_1 = \lambda S_2 \, ,
~~~~~{\cLd}_1 S_2 = -\lambda S_1 \, ,
\end{eqnarray}
where $\lambda = \sqrt{\ell(\ell +1)}$ is a separation constant.
All functions $S_0(\theta,\varphi)$, $S_1(\theta,\varphi)$ and
$S_2(\theta,\varphi)$ are, respectively, spin-weighted spherical
harmonics $_pY_{\ell m}(\theta,\varphi)$ with spin-weight $p = 1,
0, -1$, satisfying the following equation \cite{GMNRS}
\begin{eqnarray}
&&\Big[\pdta +\coa \pta +\frac{1}{\sda}\pdvi
+\frac{2ip\cta}{\sda}\pvi \nn \\
&&~~~~ - p^2\cot^2\theta +p +(\ell -p)(\ell +p +1)\Big]
{_pY}_{\ell m}(\theta,\varphi) = 0 \, .
\end{eqnarray}

As to the Hawking radiation, we should be concerned about the asymptotic
behaviors of the radial part of Eq. (\ref{sepa}) in the vicinity of the
event horizon. Because the Vaidya-de Sitter spacetime is spherically
symmetric, we can introduce as a working ansatz the following GTCT as
did in the Paper I
\begin{equation}
r_* = r +\frac{1}{2\kappa}\ln[r -r_H(v)] \, ,
~~v_* = v -v_0 \, ,\label{trans}
\end{equation}
where $r_H = r(v)$ is the location of the event horizon, and $\ka$ is an
adjustable parameter and is unchanged under the tortoise transformation.
The parameter $v_0$ is an arbitrary constant which characterizes the initial
instant of the hole. From formula (\ref{trans}), we can deduce some useful
relations for the derivatives as follows:
$$ \pr = \spr +\frac{1}{\tkr}\spr \, ,
~~\pv = \spv -\frac{\drH}{\tkr}\spr \, . $$
The quantities $\drH = \p r_H/\p v$ is the rate of the event horizon varying
in time $v$. It describes the evolution of the black hole event horizon in
the time, which reflects the presence of quantum ergosphere near the event
horizon.

Now let us consider first the asymptotic behaviors of Eq. (\ref{sepa}) near
the event horizon. Under the transformations (\ref{trans}), Eq. (\ref{sepa})
can be reduced to the following forms
\begin{eqnarray}
&&\spr R_1 = 0 \, ,
~~~~~~ 2r_H^2(\drH-G) \spr R_0 = 0 \, , \nn\\
&&2r_H^2\spr R_2 = 0 \, ,
~~~~~~ (\drH-G)\spr R_1 = 0 \, , \label{foe}
\end{eqnarray}
after being taken the $r \rightarrow r_H(v_0)$ and $v \rightarrow v_0$ limits.

From Eq. (\ref{foe}), we know that $R_1(r_*)$ and $R_2(r_*)$ are regular
on the event horizon,
\begin{equation}
\spr R_1 = \spr R_2 = 0 \, , \label{rela}
\end{equation}
thus a reasonable solution to Eq. (\ref{foe}) is that the derivatives
$\spr R_0$ does not vanish. The sole possibility we are left for the
existence of a nontrial solution of $R_0$ is (as for $r_H \not = 0$)
\begin{equation}
2G(r_H) -2\drH = 0 \, , \label{loca}
\end{equation}
which determines the location of the event horizon. It is interesting
to note that the event horizon equation (\ref{loca}) coincides with that
inferred from the null surface equation $g^{ij}\partial_i F\partial_j F
= 0$. Because $r_H$ depends on time $v$, the location of the event horizon
and the shape of the black hole change with time.

\section{Hawking Temperature}

In the preceding section, we have deduced the event horizon equation
from the limiting form of the separated radial part of the first-order
Maxwell equations. Applying a similar procedure to its second-order
forms, we can derive the Hawking temperature and the thermal radiation
spectrum. A straightforward calculation gives the second-order radial
equations
\begin{eqnarray}
&&2r^2\Big[G\pdr +\pdvr  +2(G_{,r}+\frac{G}{r})\pr \nn\\
&&~~~~ +\frac{2}{r}\pv+\frac{2G_{,r}}{r} +G_{,rr}\Big]R_1
-\lambda^2 R_1 = 0 \, ,\label{wzero}\\
&&2r^2\Big(G\pdr +\pdvr +G_{,r}\pr\Big)R_1 -\lambda^2 R_1 = 0 \, ,
\label{wone}\\
&&2r^2\Big(G\pdr +\pdvr +\frac{2G}{r}\pr\Big)R_2 -\lambda^2 R_2 = 0 \, .
\label{wtwo}
\end{eqnarray}

Given the GTCT in Eq. (\ref{trans}) and after some tedious calculations,
Eqs. (\ref{wzero}-\ref{wtwo}) have the following limiting forms near the
event horizon $r = r_H$
\begin{eqnarray}
&&\Big\{\Big[\frac{A}{2\ka} +4G(r_H) -2\drH\Big]\spdr +2\spdvr
+\Big[-A +4G_{,r}(r_H)  \nn\\
&&~~+\frac{4G(r_H)-4\drH}{r_H}\Big]\spr\Big\} R_0
=\Big\{\Big(\frac{A}{2\ka} -2\drH\Big)\spdr +2\spdvr \nn\\
&&~~~~ +\Big[-A +4G_{,r}(r_H)\Big]\spr \Big\}R_0 = 0 \, ,\label{ra0}\\
&&\Big\{\Big[\frac{A}{2\ka} +4G(r_H) -2\drH\Big]\spdr +2\spdvr
+[-A +2G_{,r}(r_H)]\spr\Big\} R_1 \nn\\
&&~~ =\Big[\Big(\frac{A}{2\ka} -2\drH\Big)\spdr +2\spdvr \Big]R_1 = 0 \, ,
\label{ra1} \\
&&\Big\{\Big[\frac{A}{2\ka} +4G(r_H) -2\drH\Big]\spdr +2\spdvr
+\Big[-A +\frac{4G(r_H)}{r_H}\Big]\spr\Big\} R_2 \nn\\
&&~~ =\Big[\Big(\frac{A}{2\ka} -2\drH\Big)\spdr +2\spdvr \Big]R_2 = 0 \, ,
\label{ra2}
\end{eqnarray}
when $r$ approaches $r_H(v_0)$ and $v$ goes to $v_0$. In the above, we
have used relations $2G(r_H) = 2\drH$ and $\spr R_1 = \spr R_2 = 0$.
The coefficient $A$ is an infinite form of $0/0$-type with a finite
result treated by using of the L' H\^{o}spital rule,
$$A = \lim_{r \rightarrow r_H(v_0)}\frac{2G -2\drH}{r -r_H}
= 2G_{,r}(r_H) \, .$$

In order to recast Eqs. (\ref{ra0}), (\ref{ra1}) and (\ref{ra2})
into a standard wave equation near the event horizon, we select the
adjustable parameter $\ka$ in them such that it satisfies
\begin{equation}
\frac{A}{2\ka} +2G(r_H) = \frac{G_{,r}(r_H)}{\ka} +2\drH \equiv 1 \, ,
\end{equation}
which means the \lq\lq surface gravity" of the horizon is
\begin{equation}
\ka =\frac{G_{,r}(r_H)}{1-2G(r_H)}= \frac{G_{,r}(r_H)}{1-2\drH} \, ,
\label{temp}
\end{equation}
where we have used Eq. (\ref{loca}).

With such a parameter adjustment, these wave equations
(\ref{ra0}-\ref{ra2}) can be reduced to
\begin{eqnarray}
&&\Big(\spdr +2\spdvr +2C \spr\Big)R_0 = 0 \, , \label{wr0} \\
&&\Big(\spdr +2\spdvr \Big)R_1 = 0 \, ,  \label{wr1} \\
&&\Big(\spdr +2\spdvr \Big)R_2 = 0 \, ,  \label{wr2}
\end{eqnarray}
where $C = G_{,r}(r_H)$. Eqs. (\ref{wr0}-\ref{wr2}) have a standard
form of wave equation in the vicinity of the event horizon. We point
out that the above parameter adjustment is an important step in our
discussions.

\section{Thermal Radiation Spectrum}

Combining Eqs. (\ref{wr1}, \ref{wr2}) with $\spr R_1 = \spr R_2 = 0$, we
know that $R_1$ and $R_2$ are independent of $r_*$ near the event horizon.
The solutions $R_1 \sim e^{-i\omega v_*}$ and $R_2 \sim e^{-i\omega v_*}$
indicate that Hawking radiation does not exist for $\Phi_1$ and $\Phi_2$.

Now separating variables to Eq. (\ref{wr0}) as $R_0 = R_0(r_*)e^{-i\omega
v_*}$, one gets
\begin{equation}
 R_0^{\prime\prime} = 2(i\omega -C)R_2^{\prime} \, ,
~~~~  R_0 = c_1 e^{2(i\omega -C)r_*} +c_2 \, .
\end{equation}
The incoming wave solution and the outgoing wave solution to Eq. (\ref{wr0})
are, respectively,
\begin{eqnarray}
&&R_0^{\rm in} \sim e^{-i\omega v_*} \, , \nn\\
&&R_0^{\rm out} \sim e^{-i\omega v_*}
e^{2(i\omega -C)r_*} \, ,~~~~~~~ (r > r_H) \, .
\end{eqnarray}
Near the event horizon, we have $r_* \sim \frac{1}{2\kappa}\ln (r - r_H)$.
Clearly, the outgoing wave $R_0^{\rm out}(r > r_H)$ is not analytic at the
event horizon $r = r_H$, but can be analytically extended from the exterior
of the hole into the interior of the hole by the lower complex $r$-plane
$$ (r -r_H) \rightarrow (r_H -r)e^{-i\pi}$$
to
\begin{equation}
\widetilde{R_0^{\rm out}} = e^{-i\omega v_* }e^{2(i\omega -C)r_*}
e^{i\pi C/\ka}e^{\pi\omega/\ka} \, ,~~~~~~(r < r_H) \, .
\end{equation}

Following the method of Damour-Ruffini-Sannan's \cite{DRS}, the relative
scattering probability of the outgoing wave at the event horizon horizon
and the thermal radiation spectrum of photons from the event horizon of
the black hole are easily obtained
\begin{equation}
\Big|\frac{R_0^{\rm out}}{\widetilde{R_0^{\rm out}}}\Big|^2
= e^{-2\pi\omega/\ka} \, ,
~~~~ \langle {\cal N}_{\omega} \rangle \sim
\frac{1}{e^{\omega/T} -1} \, , \label{sptr}
\end{equation}
in which $m$ is the azimuthal quantum number, and the Hawking
temperature is
\begin{equation}
 T = \frac{\ka}{2\pi} = \frac{1}{4\pi r_H} \cdot \frac{M r_H
-\Lambda r_H^4/3 }{M r_H -\Lambda r_H^4/6 } \, .
\end{equation}
The thermal radiation spectrum (\ref{sptr}) due to the Bose-Einstein
statistics of photons shows that the black hole emits radiation just
like a black body emitting scalar particles. The temperature depends
on the time and is consistent with that derived from the investigation
of the thermal radiation of Dirac particles in a Vaidya-de Sitter black
hole \cite{WC} with a vanishing electric charge ($Q = 0$).

\section{Conclusions}

In this paper, we have studied the Hawking radiation of photons in a
Vaidya-de Sitter black hole. The location and the temperature of the
event horizon given by Equations (\ref{loca}) and (\ref{temp}),
respectively, depend on the advanced time $v$. They can recover the
well-known results previously derived in the discussion on the Hawking
evaporation of Klein-Gordon and Dirac particles in the same spacetime.
Eq. (\ref{sptr}) shows that the thermal radiation spectra of photons
have the same form as that of Klein-Gordon particles in a Vaidya-type
black hole with a cosmological constant $\Lambda$.

In summary, we have dealt with the asymptotic behaviors of the separated
radial equations near the event horizon, not only its first-order form but
also its second-order form. We find that the limiting form of its first-order
equations puts very strong restrict on the Hawking effect, that is, not all
components of Maxwell complex scalars but $\phi_0$ displays the property of
thermal radiation. As is revealed in Paper I, we argued that this asymmetry
of Hawking radiation for different components of Maxwell fields probably
stem from the asymmetry of spacetimes in the advanced Eddington-Finkelstein
coordinate. We think this is a common character shared by the thermal radiation
of particles with higher spins in any Vaidya-type spherically symmetric black
hole.

\section*{Acknowledgment}

This work was supported partially by the NSFC in China.

\end{document}